# Electromagnetic Waves Propagation along Tangentially Magnetised Bihyrotropic Layer (with Example of Spin Waves in Ferrite Plate)

Edwin H. Lock and Sergey V. Gerus

(Kotel'nikov Instutute of Radio Engineering and Electronics (Fryazino branch), Russian Academy of Sciences, Fryazino, Moscow region, Russia)

Analytically, without magnetostatic approximation, the problem of electromagnetic wave propagation along arbitrary direction in a tangentially magnetized bihyrotropic layer has been solved. It is found that one can bring the Maxwell equations for this problem to the fourth order differential equation and the obtained biquadratic characteristic equation determines two different wave numbers $k_{x21}$ and $k_{x22}$ describing the wave distribution over the layer's thickness. The dispersion equation describing wave propagation in the bihyrotropic layer was obtained for the case of real $k_{x21}$ and $k_{x22}$ values. It is shown that in a ferrite plate, which is a special case of a bihyrotropic layer, three types of wave distribution over the plate's thickness can take place: surface-surface (when $k_{x21}$ and $k_{x22}$ are real numbers), volume-surface ($k_{x21}$ is imaginary and $k_{x22}$ is real) and volume-volume distribution ($k_{x21}$ and $k_{x22}$ are imaginary numbers). Characteristics of the surface spin wave in ferrite plate are investigated. It is found that dependences of the wave numbers $k_{x21}$ and $k_{x22}$ on the wave vector orientation are significantly different from the similar magnetostatic dependence for a large part of the wave spectrum.

## 1. Introduction

As a result of dynamical development of magnonics in the last decade, there are a lot of new data about spin waves characteristics and physical effects that can be realized by means of these waves in structures based on ferrites and antiferromagnetics [1 - 10]. It is well known that spin waves (SW) with wave numbers ~ $0 < k < 10^4$ cm$^{-1}$ are still usually described in magnetostatic approximation with assumption that spin wave number $k \gg k_0 \equiv \omega/c$ (here $\omega$ is cyclic frequency of SW and $c$ – light velocity) and, therefore one can use the magnetostatic equations to describe these waves, that is, to neglect terms containing multipliers $\omega/c$ in Maxwell equations. Such a way of SW description has been proposed in [11], which is still the most cited theoretical work devoted to SW. Due to mathematical simplicity, the results obtained in [11] have been used for more than sixty years in calculations of SW characteristics and in the design of various spin electronics devices [12 - 14]. SW are often named as magnetostatic waves (MSW) because of the use of magnetostatic approximation in their description [11].

At the same time, approximately in 80's of the twentieth century, researchers of SW had questions, the answers to which could not be found within description of SW in magnetostatic approximation. As a result, articles with studies of SW properties without magnetostatic approximation began to appear [13, 15 - 28]. The change of SW characteristics for small wave numbers $k$ was studied in some of these papers [15 - 23], including studies [18, 19, 22], where it was studied the mechanism of radiation arising from nonuniformly magnetized ferrite-dielectric structure during propagation of surface MSW at $k \to k_0$. The influence of dielectric permittivity of the medium surrounding ferrite layer on the SW dispersion dependence was investigated in [20]. It was also shown [25], that calculation of SW's Poynting vector in magnetostatic approximation by the formula $\mathbf{P} = -\omega \mathrm{Re}(i\Psi^* \mathbf{B})/8\pi$ are not correct, therefore one should calculate the Poynting vector and the power flux of SW from the known formula $\mathbf{P} = c\,\mathrm{Re}[\mathbf{EH}^*]/8\pi$ [1], finding the electric microwave field $\mathbf{E}$ from the first Maxwell equation[2]. Calculations of vector lines distribution for microwave field of surface SW [26 - 28] allowed us to understand mechanism of the wave propagation. In particular, it was found that vector lines of microwave magnetic induction form two rows of vortices localized near opposite surfaces of ferrite plate, and the vector lines of adjacent vortices always direct oppositely. The boundary between these rows of vortices is a plane (lying inside ferrite plate) on which the amplitude of microwave electric field of SW is zero. Thus, surface SW may be considered as magnetic induction vortices propagating in time and space along ferrite plate in various ferrite structures.

---

[1] In the mentioned formulas $\mathbf{E}$ and $\mathbf{H}$ are microwave electric and magnetic field vectors, $\mathbf{B}$ is magnetic induction vector and $\Psi$ –magnetostatic potential of SW.
[2] See the first equation in (4) below.

It should be noted that in all of mentioned studies, performed without magnetostatic approximation, the SW characteristics were investigated only for the case where the vectors of group and phase velocities of SW are collinear[3] (i.e., when the wave propagates perpendicular to the external magnetic field direction or along it). It is obvious that for further development of magnonics, it would be important to find an analytical solution of the problem of spin wave propagation in an arbitrary direction based on Maxwell's equations (without magnetostatic approximation). The solution of this problem would lead the description of SW to a qualitatively new level and would finally allow one not only to make accurate calculations of characteristics for SW with a non-collinear orientation of the wave vector and group velocity vector but also to calculate, for the first time, the Poynting vector, direction and density of energy flux and the structure of magnetic and electric microwave field vector lines for such waves. Moreover, the following consideration will show that solution of this problem will result in the description of fundamentally new properties of SW.

## 2. Problem statement.

It will be shown below that mathematics makes it possible to solve analytically the system of Maxwell equations (without magnetostatic approximation) and find the dispersion equation for electromagnetic waves propagating in an arbitrary direction in tangentially magnetized bihyrotropic layer of thickness $s$, (Fig. 1), which is characterized by dielectric and magnetic permeabilities described by second rank Hermite tensors $\overleftrightarrow{\varepsilon}_2$ and $\overleftrightarrow{\mu}_2$

---

[3] The only exception are papers [16, 23], which describe SW propagation in an arbitrary direction; however, as will be shown below, the results presented in these papers are not correct.

$$\overleftrightarrow{\mu_2} = \begin{vmatrix} \mu & i\nu & 0 \\ -i\nu & \mu & 0 \\ 0 & 0 & \mu_{zz} \end{vmatrix}, \qquad (1)$$

$$\overleftrightarrow{\varepsilon_2} = \begin{vmatrix} \varepsilon & ig & 0 \\ -ig & \varepsilon & 0 \\ 0 & 0 & \varepsilon_{zz} \end{vmatrix}. \qquad (2)$$

Note that the mathematical description of this problem is rather cumbersome, so we will be forced to skip the intermediate calculations below, and for a compact writing of obtained results we will introduce some notations for the different intermediate quantities.

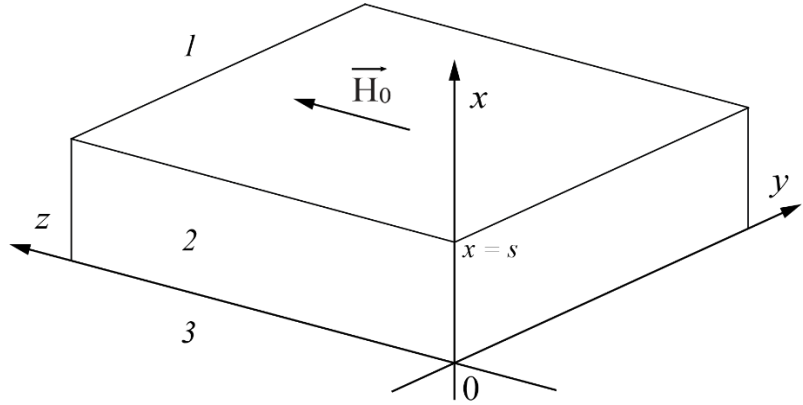

Fig. 1. Geometry of the problem: *1* and *3* - vacuum half-spaces, *2* - bihyrotropic layer (in particular case – ferrite plate) of thickness *s*.

Since the results obtained below may be useful for researchers of electromagnetic waves in different media – gyrotropic layers of ferrite, antiferromagnetic or plasma (which are special cases of bihyrotropic media and have either tensor $\overleftrightarrow{\varepsilon_2}$ or tensor $\overleftrightarrow{\mu_2}$ corresponding to expressions (1) and (2)), all mathematical deductions and formulas will be presented for the general case of wave propagation in bihyrotropic layer. At the same time, to avoid looking rather abstract, the obtained formulas will be used to calculate (as an example) the characteristics of electromagnetic waves propagating in a ferrite plate, for which the diagonal and non-diagonal components of the tensor are described by expressions [13].

$$\mu = 1 + \frac{\omega_M \omega_H}{\omega_H^2 - \omega^2}, \quad \nu = \frac{\omega_M \omega}{\omega_H^2 - \omega^2} \qquad (3)$$

where $\omega_H = \gamma H_0$, $\omega_M = 4\pi\gamma M_0$, $\gamma$ is gyromagnetic constant, $4\pi M_0$ – ferrite saturation magnetization, $f = \omega/2\pi$ – electromagnetic wave frequency. The frequency dependences of the diagonal and non-diagonal components of the tensor $\overleftrightarrow{\varepsilon_2}$ will not be detailed in this paper because the consideration below is valid for any form of these dependences, including the case where dielectric permittivity of layer 2 in Fig. 1 is a scalar quantity.

## 3. Equations describing electromagnetic wave propagation in a tangentially magnetised bihyrotropic layer.

Consider an infinite plane bihyrotropic layer 2 of thickness s, surrounded by vacuum half-spaces 1 and 3 (Fig. 1). To characterize the electromagnetic fields in media 1 – 3, let us associate with them the corresponding indices $j = 1$, 2 or 3. Layer 2 is magnetized to saturation by a tangential uniform magnetic field $\mathbf{H_0}$ and is characterized by dielectric and magnetic permeability tensors $\overleftrightarrow{\varepsilon_2}$ and $\overleftrightarrow{\mu_2}$ according to expressions (1) and (2). The half-spaces 1 and 3 have scalar relative dielectric and magnetic permittivities $\varepsilon_1$, $\mu_1$ и $\varepsilon_3$, $\mu_3$.

An electromagnetic field with frequency $\omega$, propagating in the plane of a bihyrotropic layer and changing in time according to the harmonic law $\sim \exp(i\omega t)$, must satisfy to the system of Maxwell equations[4] for complex amplitudes in each medium

$$\begin{cases} \operatorname{rot} \mathbf{E_j} + i\omega \mathbf{B_j} / c = 0 \\ \operatorname{div} \mathbf{B_j} = 0 \\ \operatorname{rot} \mathbf{H_j} - i\omega \mathbf{D_j} / c = 0 \\ \operatorname{div} \mathbf{D_j} = 0 \end{cases} \qquad (4)$$

---

[4] Exchange interaction will not be taken into account in this study.

where $\mathbf{E_j}$, $\mathbf{H_j}$ and $\mathbf{D_j}$, $\mathbf{B_j}$ are the complex amplitudes of the vectors of microwave electric and magnetic field strengths and of electric and magnetic induction, which are related by the formulas

$$\mathbf{D_j} = \overrightarrow{\varepsilon_j} \mathbf{E_j}, \quad \mathbf{B_j} = \overrightarrow{\mu_j} \mathbf{H_j} \tag{5}$$

Note here that in the previous papers in which attempts have been made to solve this problem (see, for example, [16, 23]), to describe wave propagation in a ferrite medium it was immediately proposed to find a solution of system (4) in the form of a plane wave of type $\sim \exp(-ik_x x - ik_y y - ik_z z)$. We think that from a mathematical point of view, this approach is wrong and will not allow one to find a general solution of Maxwell equations (4). In a mathematically correct approach, the dependence of the wave on the coordinate $x$ (normal to the bihyrotropic layer) should be found as a result of solving differential equations obtained by simplifying system (4) in accordance with the geometry of the problem. Therefore, we should look for solutions of system (4) in the form of homogeneous plane wave propagating in the layer plane $yz$ and characterized by an arbitrary wave vector $\mathbf{k}$. That is, in contrast to [16, 23], we leave arbitrary dependence of the field components $\mathbf{E_j}$ and $\mathbf{H_j}$ on $x$-coordinate and consider that these components change in the layer plane, as well as in time, according to the harmonic law in accordance with expressions

$$\mathbf{E_j} = \mathbf{e_j}(x)\exp(-i\mathbf{kr}) \text{ or } E_{xj,yj,zj} = e_{xj,yj,zj}(x)\exp(-ik_y y - ik_z z), \tag{6}$$

$$\mathbf{H_j} = \mathbf{h_j}(x)\exp(-i\mathbf{kr}) \text{ or } H_{xj,yj,zj} = h_{xj,yj,zj}(x)\exp(-ik_y y - ik_z z). \tag{7}$$

Besides the Cartesian coordinate system $\Sigma_D = \{x; y; z\}$, we also introduce here the corresponding polar (cylindrical) coordinate system $\Sigma_P = \{x; r; \varphi\}$, in which the angles $\varphi$ are counted from the $y$ axis, and the counterclockwise direction is taken as the positive direction of the angles. The coordinates of the systems $\Sigma_P$ and $\Sigma_D$ are related by the formulas $y = r \cos\varphi$, $z = r \sin\varphi$. Obviously, the wave vector modulus $k$ and its components $k_y$ and $k_z$ are also related by expressions $k_y = k \cos\varphi$, $k_z = k \sin\varphi$ and $k^2 = k_y^2 + k_z^2$.

Substituting expressions (6) and (7) into (5), and (5) into (4), and solving system (4) for the bihyrotropic medium *2* by analogy with [29, 30], we obtain a system of two equations containing only *x*-dependent amplitudes $e_{z2}$ and $h_{z2}$ of electromagnetic field components $E_{z2}$ and $H_{z2}$:

$$\begin{cases} \dfrac{1}{k_0^2}\dfrac{\partial^2 e_{z2}}{\partial x^2} - F_v e_{z2} - i\mu_{zz} F_{vg} h_{z2} = 0 \\ \dfrac{1}{k_0^2}\dfrac{\partial^2 h_{z2}}{\partial x^2} - F_g h_{z2} + i\varepsilon_{zz} F_{vg} e_{z2} = 0 \end{cases}, \tag{8}$$

where the dimensionless functions $F_v$, $F_g$ and $F_{vg}$ have the next form in the coordinate systems $\Sigma_D$ and $\Sigma_P$

$$F_v = \frac{k_y^2}{k_0^2} + \frac{\varepsilon_{zz}}{\varepsilon}\frac{k_z^2}{k_0^2} - \frac{\varepsilon_{zz}}{\mu}(\mu^2 - \nu^2) = \frac{k^2}{k_0^2}\left(\cos^2\varphi + \frac{\varepsilon_{zz}}{\varepsilon}\sin^2\varphi\right) - \varepsilon_{zz}\mu_\perp, \tag{9}$$

$$F_g = \frac{k_y^2}{k_0^2} + \frac{\mu_{zz}}{\mu}\frac{k_z^2}{k_0^2} - \frac{\mu_{zz}}{\varepsilon}(\varepsilon^2 - g^2) = \frac{k^2}{k_0^2}\left(\cos^2\varphi + \frac{\mu_{zz}}{\mu}\sin^2\varphi\right) - \mu_{zz}\varepsilon_\perp, \tag{10}$$

$$F_{vg} = \frac{k_z}{k_0}\left(\frac{g}{\varepsilon} + \frac{\nu}{\mu}\right) = \frac{k}{k_0}\sin\varphi\left(\frac{g}{\varepsilon} + \frac{\nu}{\mu}\right), \tag{11}$$

and the following notations are also used

$$\mu_\perp = (\mu^2 - \nu^2)/\mu, \tag{12}$$

$$\varepsilon_\perp = (\varepsilon^2 - g^2)/\varepsilon. \tag{13}$$

Note here that both non-diagonal components *v* and *g* of the tensors $\overleftrightarrow{\varepsilon}_2$ и $\overleftrightarrow{\mu}_2$ enter only into $F_{vg}$ function, whereas the *v* component enters only into $F_v$ function and the *g* component - only into $F_g$ function (that explains the use of introduced notations).

Finding the value $h_{z2}$ from the first equation of system (8) and substituting it into the second equation, we obtain the following differential equation for the amplitude $e_{z2}$

$$\frac{\partial^4 e_{z2}}{\partial x^4} + 2\eta\frac{\partial^2 e_{z2}}{\partial x^2} + \alpha e_{z2} = 0, \tag{14}$$

where

$$\eta = -k_0^2(F_v + F_g)/2 \tag{15}$$

$$\alpha = k_0^4 F_v F_g - \mu_{zz}\varepsilon_{zz}k_0^4 F_{vg}^2. \tag{16}$$

The following characteristic equation, corresponding to equation (14), determines the values of wave number $k_{x2}$ inside the bihyrotropic layer

$$k_{x2}^4 + 2\eta k_{x2}^2 + \alpha = 0. \tag{17}$$

Using expressions (15) and (16), it is easy to show that discriminant of equation (17) can take only positive values:

$$\eta^2 - \alpha = k_0^4 \left(F_v + F_g\right)^2 / 4 - k_0^4 \left(F_v F_g - \mu_{zz}\varepsilon_{zz} F_{vg}^2\right) =$$
$$= k_0^4 \left[\left(F_v - F_g\right)^2 / 4 + \mu_{zz}\varepsilon_{zz} F_{vg}^2\right] > 0. \tag{18}$$

The characteristic equation (17) has four roots defined by the expression

$$k_{x2}^2 = -\eta \pm \sqrt{\eta^2 - \alpha} = \frac{k_0^2}{2}\left(F_v + F_g \pm \sqrt{\left(F_v - F_g\right)^2 + 4\mu_{zz}\varepsilon_{zz}F_{vg}^2}\right) \tag{19}$$

and all roots are simple (not multiples):

$$k_{x21} = \sqrt{-\eta - \sqrt{\eta^2 - \alpha}} = k_0\sqrt{\frac{F_v + F_g}{2} - \frac{1}{2}\sqrt{\left(F_v - F_g\right)^2 + 4\mu_{zz}\varepsilon_{zz}F_{vg}^2}}, \tag{20}$$

$$k_{x22} = \sqrt{-\eta + \sqrt{\eta^2 - \alpha}} = k_0\sqrt{\frac{F_v + F_g}{2} + \frac{1}{2}\sqrt{\left(F_v - F_g\right)^2 + 4\mu_{zz}\varepsilon_{zz}F_{vg}^2}}, \tag{21}$$

$$k_{x23} = -k_{x21}, \tag{22}$$

$$k_{x24} = -k_{x22}. \tag{23}$$

**4. Solutions describing electromagnetic waves in a bihyrotropic layer.**

To determine solutions of the differential equation (14), we need to find out what values the roots $k_{x21} - k_{x24}$ can take. First, note that the roots $k_{x21} - k_{x24}$ cannot be complex numbers (since according to (18) always $\eta^2 - \alpha > 0$), but can take only real or imaginary values, depending on the sign of radicands in (20) and (21). As can be seen from these expressions, if $\alpha < 0$, then $|\eta|$ is always smaller than value $\sqrt{\eta^2 - \alpha}$, the sign before which determines the radicand sign; in this case $k_{x21}$ is always imaginary, while $k_{x22}$ is always real (for any sign of $\eta$). If $\alpha > 0$, conversely, it is always $|\eta| > \sqrt{\eta^2 - \alpha}$, and both $k_{x21}$ and $k_{x22}$ have imaginary values when $\eta > 0$, and real values when $\eta < 0$.

Thus, following the conditions formulated above and using expressions (15), (16) and (9) – (11), one can plot the boundary surfaces for certain parameters of bihyrotropic layer in coordinate spaces $\{k_y, k_z, f\}$ or $\{k, \varphi, f\}$. The equations for these surfaces according to (15) and (16) can be written in the next form

$$\eta = 0 \text{ or } F_v + F_g = 0 \qquad (24)$$

$$\alpha = 0 \text{ or } F_v F_g - \mu_{zz} \varepsilon_{zz} F_{vg}^2 = 0, \qquad (25)$$

The boundary surfaces have a simple physical meaning: intersecting certain dispersion surface $f(k_y, k_z)$ of electromagnetic waves[5], the boundary surfaces will separate on it areas with real and imaginary values of the roots $k_{x21} - k_{x24}$.

Note first of all that expression (25) is identical to the dispersion equation for electromagnetic waves in an unbounded bihyrotropic medium (see relations (20) – (23) in [30]), if this equation would be simplified to the two-dimensional case, by equating to zero the wave number for one of coordinates normal to vector **H₀** (for example, choosing $k_x = 0$ in relations (20) – (23) in [30]).

To imagine clearly the surfaces $\alpha = 0$ and $\eta = 0$, let us make calculations for the case where layer *2* in Fig. 1 is a ferrite plate (which is a special case of a bihyrotropic layer) having saturation magnetization $4\pi M_0 = 1750$ Gc and dielectric permittivity $\varepsilon_2 = 15$. The value of homogeneous magnetic field $H_0$ magnetizing the plate to saturation was equal to 300 Oe in calculations.

Note at once that for the case with ferrite plate the value $\alpha$ changes sign when the frequency changes from value $f < f_\perp$ to value $f > f_\perp$ (since according to (3) and (12) we have $\mu = 0$ and $\mu_\perp \to \infty$ at $f = f_\perp$).

Now, to find out in which regions of space $\{k_y, k_z, f\}$ the roots $k_{x21}$ и $k_{x22}$ take real values, and in which regions they take imaginary values, we calculate and graph in this space the surfaces $\alpha = 0$, $\eta = 0$ and the plane $f = f_\perp$. The spatial regions bounded by these surfaces and cross sections of these surfaces with planes $k_y = 0$, $k_z = 0$ and $f = 7000$ MHz are shown in Fig. 2.

---

[5] For example, a dispersion surface for some type of spin waves propagating in a ferrite plate.

The meaning of 3-D chart in Fig.2 is that we know what distribution over the thickness of tangentially magnetized ferrite layer (with an arbitrary thickness!) the wave will have if its dispersion surface will be in a certain region of space $\{k_y, k_z, f\}$. That is, we know about these space regions now, based only on the properties of differential equation (14), although we have not yet got the dispersion equation of the wave!

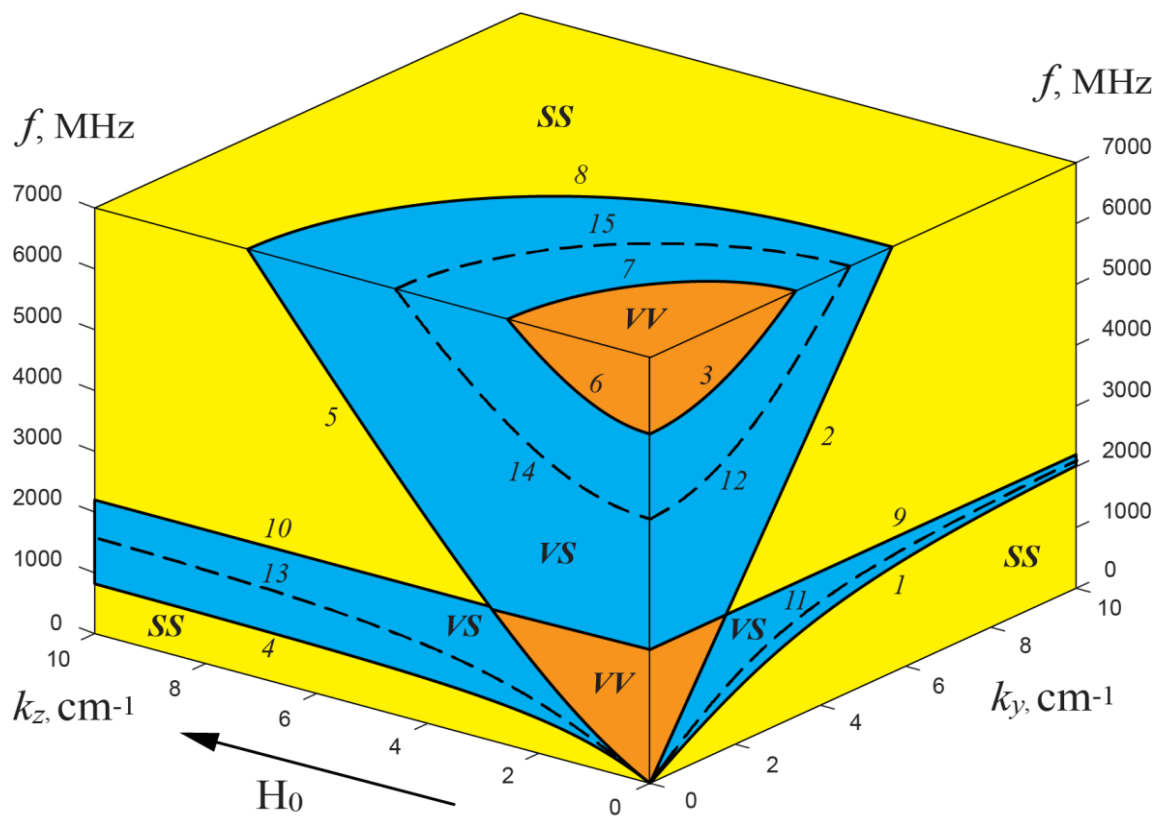

Fig. 2. Spatial regions *SS*, *VS* and *VV* defining the character of wave distribution in a ferrite plate cross section. The boundaries of the *SS*, *VS* and *VV* regions are defined by the united surface $\alpha = 0$ (which itself includes several surfaces) and the plane $f = f_\perp$. Curves 1 - 3, 4 - 6 and 7 - 8 correspond to sections of the surface $\alpha = 0$ by planes $k_y = 0$, $k_z = 0$ and $f = 7000$ MHz respectively. Lines 9 and 10 are sections of plane $f = f_\perp$ by planes $k_y = 0$ and $k_z = 0$ respectively. Dashed curves 11 - 12, 13 - 14 and 15 are sections of the surface $\eta = 0$ by planes $k_y = 0$, $k_z = 0$ and $f = 7000$ MHz respectively.

Analysing Fig. 2, we note that the most voluminous in Fig. 2 are the SS-regions highlighted by yellow, where α > 0 and η < 0. The parts of the wave dispersion surface located in the SS-regions will describe solutions with real values of roots $k_{x21}$ и $k_{x22}$, to which corresponds the general solution of differential equation (14) in the form

$$e_{z2} = A\exp(k_{x21}x) + B\exp(-k_{x21}x) + C\exp(k_{x22}x) + D\exp(-k_{x22}x). \qquad (26)$$

That is, the wave distribution over the ferrite plate thickness for these parts of the dispersion surface, will be described only by exponential functions and such a wave can be conventionally called a surface-surface or SS-wave.

The smallest space in Figure 2 is occupied by VV-regions, highlighted by orange, where α > 0 and η > 0. The parts of the wave dispersion surface located in VV-regions will describe solutions with imaginary values of roots $k_{x21}$ and $k_{x22}$, to which corresponds a general solution of the differential equation (22) in the form of

$$e_{z2} = A\cos(k_{x21}x) + B\sin(-k_{x21}x) + C\cos(k_{x22}x) + D\sin(-k_{x22}x). \qquad (27)$$

Thus, the wave distribution over the ferrite plate thickness for this part of the dispersion surface, will be described only by trigonometric functions and such a wave can be conventionally called a volume-volume or VV-wave.

The relation α < 0 is valid for the VS-regions highlighted by blue in Fig. 2, and the surfaces η = 0 (their cross sections 11 – 15 are shown by dashed curves) always lie within VS-regions. As mentioned above, the value $k_{x21}$ is imaginary and the value $k_{x22}$ is real in these regions regardless of the sign of η. That is, the parts of the wave dispersion surface located in the VS-regions will describe waves corresponding to the general solution of the differential equation (22) of the form

$$e_{z2} = A\cos(k_{x21}x) + B\sin(-k_{x21}x) + C\exp(k_{x22}x) + D\exp(-k_{x22}x). \qquad (28)$$

In other words, the distribution over the ferrite plate thickness for these parts of the dispersion surface, will be described by both trigonometric and exponential functions and the wave can be conventionally called a bulk-surface or VS-wave.

It should be noted that the last case, when a general solution of the differential equation (22) has the form,

$$e_{z2} = A\exp(k_{x21}x) + B\exp(-k_{x21}x) + C\cos(k_{x22}x) + D\sin(-k_{x22}x), \qquad (29)$$

corresponding to real values of $k_{x21}$ and imaginary values of $k_{x22}$, is never realized in the ferrite plate.

Thus, the wave distribution inside the ferrite plate can vary depending on the wave parameters and on the part of dispersion or isofrequency dependence[6] this distribution can correspond, for example, to SS-wave (described by expression (26)) and on another part to VS-wave (described by expression (28)). This is an essential difference between the exact SW description and SW description in magnetostatic approximation [11], where every SW dispersion surface was characterized by the certain type of SW distribution (surface or volume).

It is also evident that there is a single-valued correspondence between the regions of space $\{k_y, k_z, f\}$ and the type of wave distribution in the ferrite layer regardless of the boundary problem we will be considering further – it can be a simple ferrite plate surrounded by vacuum half-spaces, a one-sided metallized ferrite plate, a metal-dielectric-ferrite- dielectric-metal structure, etc.

Below we obtain the dispersion equation and expressions for the microwave electromagnetic field components inside the bihyrotropic layer for SS-waves described by expression (26). Obviously, similar relations for other types of waves can be obtained by the same way.

## 5. Expressions for electromagnetic field components inside bihyrotropic layer

To obtain expressions for all microwave components of electromagnetic wave inside bihyrotropic layer, it is necessary first to find expressions for their amplitudes $e_{x2}$, $e_{y2}$, $h_{x2}$, $h_{y2}$ и $h_{z2}$.

Substituting expression (26) into the first equation of system (8), we find the expression for the amplitude $h_{z2}$

---

[6] It is well known that the dispersion and isofrequency dependences are cross sections of the dispersion surface, so all that has been said above about intersections of this surface with boundary surfaces is applicable to these dependences too.

$$h_{z2} = i\beta_1\left(A\exp(k_{x21}x) + B\exp(-k_{x21}x)\right) + i\beta_2\left(C\exp(k_{x22}x) + D\exp(-k_{x22}x)\right), \quad (30)$$

where

$$\beta_1 = \frac{1}{\mu_{zz}F_{vg}}\left(F_v - \frac{k_{x21}^2}{k_0^2}\right), \quad (31)$$

$$\beta_2 = \frac{1}{\mu_{zz}F_{vg}}\left(F_v - \frac{k_{x22}^2}{k_0^2}\right). \quad (32)$$

Substituting expressions (6) and (7) into (5) and then (5) into system (4), we obtain

$$k_y e_{z2} - k_z e_{y2} - k_0\mu h_{x2} - ik_0\nu h_{y2} = 0 \quad (33)$$

$$-ik_z e_{x2} - \frac{\partial e_{z2}}{\partial x} + k_0\nu h_{x2} + ik_0\mu h_{y2} = 0 \quad (34)$$

$$\frac{\partial e_{y2}}{\partial x} + ik_y e_{x2} + ik_0\mu_{zz} h_{z2} = 0 \quad (35)$$

$$\mu\left(\frac{\partial h_{x2}}{\partial x} - ik_y h_{y2}\right) + i\nu\left(\frac{\partial h_{y2}}{\partial x} + ik_y h_{x2}\right) - i\mu_{zz}k_z h_{z2} = 0 \quad (36)$$

$$k_y h_{z2} - k_z h_{y2} + k_0\varepsilon e_{x2} + ik_0 g e_{y2} = 0 \quad (37)$$

$$ik_z h_{x2} + \frac{\partial h_{z2}}{\partial x} + k_0 g e_{x2} + ik_0\varepsilon e_{y2} = 0 \quad (38)$$

$$\frac{\partial h_{y2}}{\partial x} + ik_y h_{x2} - ik_0\varepsilon_{zz} e_{z2} = 0 \quad (39)$$

$$\varepsilon\left(\frac{\partial e_{x2}}{\partial x} - ik_y e_{y2}\right) + ig\left(\frac{\partial e_{y2}}{\partial x} + ik_y e_{x2}\right) - ik_z\varepsilon_{zz}e_{z2} = 0 \quad (40)$$

From equations (33) and (37), we can respectively obtain the expressions

$$h_{x2} = \frac{k_y}{\mu k_0}e_{z2} - \frac{k_z}{\mu k_0}e_{y2} - i\frac{\nu}{\mu}h_{y2}, \quad (41)$$

$$e_{x2} = -\frac{k_y}{\varepsilon k_0}h_{z2} + \frac{k_z}{\varepsilon k_0}h_{y2} - i\frac{g}{\varepsilon}e_{y2}. \quad (42)$$

Substituting expressions (41) and (42) into equations (34) and (38), we obtain the following relations

$$i\frac{k_y k_z}{\varepsilon k_0^2}h_{z2} - iF_{v2}h_{y2} - F_{vg}e_{y2} + \frac{\nu k_y}{\mu k_0}e_{z2} - \frac{1}{k_0}\frac{\partial e_{z2}}{\partial x} = 0, \quad (43)$$

$$-i\frac{k_y k_z}{\mu k_0^2}e_{z2} + iF_{g2}e_{y2} - F_{vg}h_{y2} + \frac{gk_y}{\varepsilon k_0}h_{z2} - \frac{1}{k_0}\frac{\partial h_{z2}}{\partial x} = 0, \quad (44)$$

where the dimensionless functions $F_{v2}$ and $F_{g2}$ have the form

$$F_{v2} = \frac{k_z^2}{\varepsilon k_0^2} - \mu_\perp, \quad (45)$$

$$F_{g2} = \frac{k_z^2}{\mu k_0^2} - \varepsilon_\perp, \quad (46)$$

Let us multiply expression (43) and $iF_{vg}$. Then let's multiply expression (44) and $F_{v2}$. Summing the resulting expressions, we find the value $e_{y2}$

$$e_{y2} = \frac{1}{F_2}\left[a_0 e_{z2} - ia_2 h_{z2} + \frac{F_{vg}}{k_0}\frac{\partial e_{z2}}{\partial x} - i\frac{F_{v2}}{k_0}\frac{\partial h_{z2}}{\partial x}\right]. \quad (47)$$

Now let us multiply expression (43) and $iF_{g2}$. Then let's multiply expression (44) and $F_{vg}$. Summing the obtained relations, we find the value $h_{y2}$

$$h_{y2} = \frac{1}{F_2}\left[ib_0 e_{z2} + b_2 h_{z2} + \frac{F_{vg}}{k_0}\frac{\partial h_{z2}}{\partial x} + i\frac{F_{g2}}{k_0}\frac{\partial e_{z2}}{\partial x}\right]. \quad (48)$$

In expressions (47) and (48) we have introduced the following notations

$$F_2 = F_{v2}F_{g2} - F_{vg}^2, \quad (49)$$

$$a_0 = \frac{k_y k_z}{\mu k_0^2}F_{v2} - \frac{\nu k_y}{\mu k_0}F_{vg}, \quad (50)$$

$$a_2 = \frac{k_y k_z}{\varepsilon k_0^2}F_{vg} - \frac{gk_y}{\varepsilon k_0}F_{v2}, \quad (51)$$

$$b_0 = \frac{k_y k_z}{\mu k_0^2}F_{vg} - \frac{\nu k_y}{\mu k_0}F_{g2}, \quad (52)$$

$$b_2 = \frac{k_y k_z}{\varepsilon k_0^2}F_{g2} - \frac{gk_y}{\varepsilon k_0}F_{vg}. \quad (53)$$

As can be seen from (47) and (48), the amplitudes $e_{y2}$ and $h_{y2}$ are expressed only through the amplitudes $e_{z2}$, $h_{z2}$ and their x-coordinate derivatives. Substituting expressions (47) and (48) into (41) and (42), we obtain similar expressions for the amplitudes $e_{x2}$ and $h_{x2}$. To concisely record the $x$-coordinate dependencies of all amplitudes, we introduce such dimensionless functions $\Sigma_0$, $\Sigma_1$, $\Sigma_2$ and $\Sigma_3$, that the following relations are satisfied

$$e_{z2} = \Sigma_0(x), \quad \frac{\partial e_{z2}}{\partial x} = k_0 \Sigma_1(x), \quad h_{z2} = i\Sigma_2(x), \quad \frac{\partial h_{z2}}{\partial x} = ik_0 \Sigma_3(x). \quad (54)$$

Substituting formulae (26) and (30) into relations (54), we obtain the next expressions for the functions $\Sigma_0$, $\Sigma_1$, $\Sigma_2$ and $\Sigma_3$

$$\Sigma_0(x) = A\exp(k_{x21}x) + B\exp(-k_{x21}x) + C\exp(k_{x22}x) + D\exp(-k_{x22}x) \quad (55)$$

$$\Sigma_1(x) = \frac{k_{x21}}{k_0}\left(A\exp(k_{x21}x) - B\exp(-k_{x21}x)\right) +$$

$$+ \frac{k_{x22}}{k_0}\left(C\exp(k_{x22}x) - D\exp(-k_{x22}x)\right) \quad (56)$$

$$\Sigma_2(x) = \beta_1\left(A\exp(k_{x21}x) + B\exp(-k_{x21}x)\right) +$$

$$+ \beta_2\left(C\exp(k_{x22}x) + D\exp(-k_{x22}x)\right), \quad (57)$$

$$\Sigma_3(x) = \frac{k_{x21}}{k_0}\beta_1\left(A\exp(k_{x21}x) - B\exp(-k_{x21}x)\right) +$$

$$+ \frac{k_{x22}}{k_0}\beta_2\left(C\exp(k_{x22}x) - D\exp(-k_{x22}x)\right) \quad (58)$$

To explain the introduced notations, note that numerical indexes of the values $\Sigma_0$ – $\Sigma_3$ correspond to the maximum power of the wave numbers $k_{x21}$ and $k_{x22}$ in multipliers near the exponents entering in expressions (55) – (58) (the power of $k_{x21}$ and $k_{x22}$ in relations (31) and (32) for values $\beta_1$ and $\beta_2$ is taken into account too).

Substituting relations (54) - (58) into expressions (26), (30), (41), (42), (47) and (48) for amplitudes $e_{x2}$, $e_{y2}$, $h_{x2}$, $h_{y2}$ and $h_{z2}$ and then substituting obtained expressions into formulae (6) and (7), we find all components of the microwave electromagnetic field inside the bihyrotropic layer

$$E_{x2} = \frac{i}{\varepsilon F_2}\left[\frac{k_z}{k_0}\left(b_0\Sigma_0 + F_{g2}\Sigma_1 + b_2\Sigma_2 + F_{vg}\Sigma_3\right) - \frac{k_y}{k_0}F_2\Sigma_2 - \right.$$

$$-g\left(a_0\Sigma_0 + F_{vg}\Sigma_1 + a_2\Sigma_2 + F_{v2}\Sigma_3\right)\bigg]\exp(-ik_y y - ik_z z), \tag{59}$$

$$H_{x2} = \frac{1}{\mu F_2}\left[\frac{k_y}{k_0}F_2\Sigma_0 - \frac{k_z}{k_0}\left(a_0\Sigma_0 + F_{vg}\Sigma_1 + a_2\Sigma_2 + F_{v2}\Sigma_3\right) + \right.$$

$$\left. + \nu\left(b_0\Sigma_0 + F_{g2}\Sigma_1 + b_2\Sigma_2 + F_{vg}\Sigma_3\right)\right]\exp(-ik_y y - ik_z z), \tag{60}$$

$$E_{y2} = \frac{1}{F_2}\left[a_0\Sigma_0 + F_{vg}\Sigma_1 + a_2\Sigma_2 + F_{v2}\Sigma_3\right]\exp(-ik_y y - ik_z z), \tag{61}$$

$$H_{y2} = \frac{i}{F_2}\left[b_0\Sigma_0 + F_{g2}\Sigma_1 + b_2\Sigma_2 + F_{vg}\Sigma_3\right]\exp(-ik_y y - ik_z z), \tag{62}$$

$$E_{z2} = \Sigma_0(x)\exp(-ik_y y - ik_z z), \tag{63}$$

$$H_{z2} = i\Sigma_2(x)\exp(-ik_y y - ik_z z), \tag{64}$$

## 6. Expressions for electromagnetic field components outside bihyrotropic layer

Let us now consider microwave fields arising outside the bihyrotropic layer in media *1* and *3* characterized by scalar dielectric and magnetic permittivities $\varepsilon_1$, $\mu_1$ and $\varepsilon_3$, $\mu_3$. Substituting solutions of the form (6) and (7) into Maxwell equations (4), we obtain instead of system (8) two independent differential equations with respect to amplitudes $e_{z1,3}$ and $h_{z1,3}$:

$$\frac{\partial^2 e_{z1,3}}{\partial x^2} - \left(k_z^2 + k_y^2 - k_0^2\varepsilon_{1,3}\mu_{1,3}\right)e_{z1,3} = 0, \tag{65}$$

$$\frac{\partial^2 h_{z1,3}}{\partial x^2} - \left(k_z^2 + k_y^2 - k_0^2\varepsilon_{1,3}\mu_{1,3}\right)h_{z1,3} = 0, \tag{66}$$

The following characteristic equation determines the solutions of equations (65) and (66)

$$k_{x1,3}^2 = k_z^2 + k_y^2 - k_0^2\varepsilon_{1,3}\mu_{1,3}. \tag{67}$$

Since the microwave fields should exponentially decay far away from the layer, the solutions of equations (65) and (66) in medium *1* will be looked for in the form

$$e_{z1} = N\exp(-k_{x1}x), \tag{68}$$

$$h_{z1} = iG\exp(-k_{x1}x), \tag{69}$$

and in medium *3* – in the form

$$e_{z3} = K\exp(k_{x3}x), \tag{70}$$

$$h_{z3} = iL\exp(k_{x3}x), \tag{71}$$

where *N*, *G*, *L* and *K* are independent coefficients.

Transforming the system of Maxwell equations (4), we express the values $e_{x1,3}$, $h_{y1,3}$ and $h_{x1,3}$ through the values $e_{z1,3}$ and $h_{z1,3}$, described by expressions (68) – (71), and then we substitute all obtained relations in (6), (7) and find expressions for microwave field components in half-spaces *1* and *3*:

$$E_{x1} = \frac{i}{q_1^2}\left(Gk_y k_0 \mu_1 - Nk_z k_{x1}\right)\exp(-k_{x1}x - ik_y y - ik_z z). \tag{72}$$

$$H_{x1} = \frac{1}{q_1^2}\left(Gk_z k_{x1} - Nk_y k_0 \varepsilon_1\right)\exp(-k_{x1}x - ik_y y - ik_z z), \tag{73}$$

$$E_{y1} = \frac{1}{q_1^2}\left(Nk_y k_z - Gk_{x1} k_0 \mu_1\right)\exp(-k_{x1}x - ik_y y - ik_z z), \tag{74}$$

$$H_{y1} = \frac{i}{q_1^2}\left(Gk_y k_z - Nk_{x1} k_0 \varepsilon_1\right)\exp(-k_{x1}x - ik_y y - ik_z z), \tag{75}$$

$$E_{z1} = N\exp(-k_{x1}x - ik_y y - ik_z z), \tag{76}$$

$$H_{z1} = iG\exp(-k_{x1}x - ik_y y - ik_z z), \tag{77}$$

$$E_{x3} = \frac{i}{q_3^2}\left(Lk_y k_0 \mu_3 + Kk_z k_{x3}\right)\exp(k_{x3}x - ik_y y - ik_z z). \tag{78}$$

$$H_{x3} = -\frac{1}{q_3^2}\left(Lk_z k_{x3} + Kk_y k_0 \varepsilon_3\right)\exp(k_{x3}x - ik_y y - ik_z z), \tag{79}$$

$$E_{y3} = \frac{1}{q_3^2}\left(Kk_y k_z + Lk_{x3} k_0 \mu_3\right)\exp(k_{x3}x - ik_y y - ik_z z), \tag{80}$$

$$H_{y3} = \frac{i}{q_3^2}\left(Lk_y k_z + Kk_{x3} k_0 \varepsilon_3\right)\exp(k_{x3}x - ik_y y - ik_z z), \tag{81}$$

$$E_{z3} = K\exp(k_{x3}x - ik_y y - ik_z z), \tag{82}$$

$$H_{z3} = iL\exp(k_{x3}x - ik_y y - ik_z z), \quad (83)$$

where the values $q_1$ and $q_3$ are described by the following expression

$$q_{1,3}^2 = k_z^2 - k_0^2 \varepsilon_{1,3} \mu_{1,3}. \quad (84)$$

## 7. Dispersion equation for electromagnetic waves in bihyrotropic layer

Let us now proceed to the derivation of the dispersion equation describing the propagation of electromagnetic waves in a bihyrotropic layer. Satisfying the boundary conditions of continuity of tangential components $E_y$, $E_z$, $H_y$ and $H_z$ at $x = 0$ and $x = s$, one can obtain the following system of eight equations for constant coefficients $A$, $B$, $C$, $D$, $G$, $N$, $K$, $L$:

$$
\begin{aligned}
&N\exp(-k_{x1}s) = \Sigma_0(s) \\
&F_2 \frac{Nk_y k_z - G\mu_1 k_{x1} k_0}{q_1^2 \exp(k_{x1}s)} = a_0 \Sigma_0(s) + F_{vg}\Sigma_1(s) + a_2\Sigma_2(s) + F_{v2}\Sigma_3(s) \\
&G\exp(-k_{x1}s) = \Sigma_2(s) \\
&F_2 \frac{Gk_y k_z - N\varepsilon_1 k_{x1} k_0}{q_1^2 \exp(k_{x1}s)} = b_0 \Sigma_0(s) + F_{g2}\Sigma_1(s) + b_2\Sigma_2(s) + F_{vg}\Sigma_3(s) \\
&K = \Sigma_0(0) \\
&F_2 \frac{Kk_y k_z + L\mu_3 k_{x3} k_0}{q_3^2} = a_0 \Sigma_0(0) + F_{vg}\Sigma_1(0) + a_2\Sigma_2(0) + F_{v2}\Sigma_3(0) \\
&L = \Sigma_2(0) \\
&F_2 \frac{Lk_y k_z + K\varepsilon_3 k_{x3} k_0}{q_3^2} = b_0 \Sigma_0(0) + F_{g2}\Sigma_1(0) + b_2\Sigma_2(0) + F_{vg}\Sigma_3(0)
\end{aligned}
\quad (85)
$$

Substituting the values $N$, $G$, $K$ and $L$ from the first, third, fifth and seventh equations into the second, fourth, sixth and eighth equations of system (85), we obtain a system of four equations for the coefficients $A$, $B$, $C$ and $D$ (entering into the values $\Sigma_0 - \Sigma_3$):

$$\left(a_0 - \frac{k_y k_z}{q_1^2} F_2\right)\Sigma_0(s) + F_{vg}\Sigma_1(s) + \left(a_2 + \frac{\mu_1 k_{x1} k_0}{q_1^2} F_2\right)\Sigma_2(s) + F_{v2}\Sigma_3(s) = 0$$

$$\left(b_0 + \frac{\varepsilon_1 k_{x1} k_0}{q_1^2} F_2\right)\Sigma_0(s) + F_{g2}\Sigma_1(s) + \left(b_2 - \frac{k_y k_z}{q_1^2} F_2\right)\Sigma_2(s) + F_{vg}\Sigma_3(s) = 0$$

$$\left(a_0 - \frac{k_y k_z}{q_3^2} F_2\right)\Sigma_0(0) + F_{vg}\Sigma_1(0) + \left(a_2 - \frac{\mu_3 k_{x3} k_0}{q_3^2} F_2\right)\Sigma_2(0) + F_{v2}\Sigma_3(0) = 0$$

$$\left(b_0 - \frac{\varepsilon_3 k_{x3} k_0}{q_3^2} F_2\right)\Sigma_0(0) + F_{g2}\Sigma_1(0) + \left(b_2 - \frac{k_y k_z}{q_3^2} F_2\right)\Sigma_2(0) + F_{vg}\Sigma_3(0) = 0$$

(86)

Substituting expressions (55) - (58) describing the values $\Sigma_0 - \Sigma_3$ into the system (86), and collecting similar summands with the same coefficients $A$, $B$, $C$ and $D$, we obtain a system of equations

$$\begin{cases} d_{11}A + d_{12}B + d_{13}C + d_{14}D = 0 \\ d_{21}A + d_{22}B + d_{23}C + d_{24}D = 0 \\ d_{31}A + d_{32}B + d_{33}C + d_{34}D = 0 \\ d_{41}A + d_{42}B + d_{43}C + d_{44}D = 0 \end{cases}, \quad (87)$$

where the matrix elements have the form

$$d_{11} = \left[a_0 - \frac{k_y k_z}{q_1^2} F_2 + \frac{k_{x21}}{k_0} F_{vg} + \beta_1\left(a_2 + \frac{\mu_1 k_{x1} k_0}{q_1^2} F_2 + \frac{k_{x21}}{k_0} F_{v2}\right)\right]\exp(k_{x21}s) \quad (88)$$

$$d_{12} = \left[a_0 - \frac{k_y k_z}{q_1^2} F_2 - \frac{k_{x21}}{k_0} F_{vg} + \beta_1\left(a_2 + \frac{\mu_1 k_{x1} k_0}{q_1^2} F_2 - \frac{k_{x21}}{k_0} F_{v2}\right)\right]\exp(-k_{x21}s) \quad (89)$$

$$d_{13} = \left[a_0 - \frac{k_y k_z}{q_1^2} F_2 + \frac{k_{x22}}{k_0} F_{vg} + \beta_2\left(a_2 + \frac{\mu_1 k_{x1} k_0}{q_1^2} F_2 + \frac{k_{x22}}{k_0} F_{v2}\right)\right]\exp(k_{x22}s) \quad (90)$$

$$d_{14} = \left[a_0 - \frac{k_y k_z}{q_1^2} F_2 - \frac{k_{x22}}{k_0} F_{vg} + \beta_2\left(a_2 + \frac{\mu_1 k_{x1} k_0}{q_1^2} F_2 - \frac{k_{x22}}{k_0} F_{v2}\right)\right]\exp(-k_{x22}s) \quad (91)$$

$$d_{21} = \left[b_0 + \frac{\varepsilon_1 k_{x1} k_0}{q_1^2} F_2 + \frac{k_{x21}}{k_0} F_{g2} + \beta_1\left(b_2 - \frac{k_y k_z}{q_1^2} F_2 + \frac{k_{x21}}{k_0} F_{vg}\right)\right]\exp(k_{x21}s) \quad (92)$$

$$d_{22} = \left[b_0 + \frac{\varepsilon_1 k_{x1} k_0}{q_1^2} F_2 - \frac{k_{x21}}{k_0} F_{g2} + \beta_1\left(b_2 - \frac{k_y k_z}{q_1^2} F_2 - \frac{k_{x21}}{k_0} F_{vg}\right)\right]\exp(-k_{x21}s) \quad (93)$$

$$d_{23} = \left[ b_0 + \frac{\varepsilon_1 k_{x1} k_0}{q_1^2} F_2 + \frac{k_{x22}}{k_0} F_{g2} + \beta_2 \left( b_2 - \frac{k_y k_z}{q_1^2} F_2 + \frac{k_{x22}}{k_0} F_{vg} \right) \right] \exp(k_{x22} s) \quad (94)$$

$$d_{24} = \left[ b_0 + \frac{\varepsilon_1 k_{x1} k_0}{q_1^2} F_2 - \frac{k_{x22}}{k_0} F_{g2} + \beta_2 \left( b_2 - \frac{k_y k_z}{q_1^2} F_2 - \frac{k_{x22}}{k_0} F_{vg} \right) \right] \exp(-k_{x22} s) \quad (95)$$

$$d_{31} = a_0 - \frac{k_y k_z}{q_3^2} F_2 + \frac{k_{x21}}{k_0} F_{vg} + \beta_1 \left( a_2 - \frac{\mu_3 k_{x3} k_0}{q_3^2} F_2 + \frac{k_{x21}}{k_0} F_{v2} \right) \quad (96)$$

$$d_{32} = a_0 - \frac{k_y k_z}{q_3^2} F_2 - \frac{k_{x21}}{k_0} F_{vg} + \beta_1 \left( a_2 - \frac{\mu_3 k_{x3} k_0}{q_3^2} F_2 - \frac{k_{x21}}{k_0} F_{v2} \right) \quad (97)$$

$$d_{33} = a_0 - \frac{k_y k_z}{q_3^2} F_2 + \frac{k_{x22}}{k_0} F_{vg} + \beta_2 \left( a_2 - \frac{\mu_3 k_{x3} k_0}{q_3^2} F_2 + \frac{k_{x22}}{k_0} F_{v2} \right) \quad (98)$$

$$d_{34} = a_0 - \frac{k_y k_z}{q_3^2} F_2 - \frac{k_{x22}}{k_0} F_{vg} + \beta_2 \left( a_2 - \frac{\mu_3 k_{x3} k_0}{q_3^2} F_2 - \frac{k_{x22}}{k_0} F_{v2} \right) \quad (99)$$

$$d_{41} = b_0 - \frac{\varepsilon_3 k_{x3} k_0}{q_3^2} F_2 + \frac{k_{x21}}{k_0} F_{g2} + \beta_1 \left( b_2 - \frac{k_y k_z}{q_3^2} F_2 + \frac{k_{x21}}{k_0} F_{vg} \right) \quad (100)$$

$$d_{42} = b_0 - \frac{\varepsilon_3 k_{x3} k_0}{q_3^2} F_2 - \frac{k_{x21}}{k_0} F_{g2} + \beta_1 \left( b_2 - \frac{k_y k_z}{q_3^2} F_2 - \frac{k_{x21}}{k_0} F_{vg} \right) \quad (101)$$

$$d_{43} = b_0 - \frac{\varepsilon_3 k_{x3} k_0}{q_3^2} F_2 + \frac{k_{x22}}{k_0} F_{g2} + \beta_2 \left( b_2 - \frac{k_y k_z}{q_3^2} F_2 + \frac{k_{x22}}{k_0} F_{vg} \right) \quad (102)$$

$$d_{44} = b_0 - \frac{\varepsilon_3 k_{x3} k_0}{q_3^2} F_2 - \frac{k_{x22}}{k_0} F_{g2} + \beta_2 \left( b_2 - \frac{k_y k_z}{q_3^2} F_2 - \frac{k_{x22}}{k_0} F_{vg} \right) \quad (103)$$

Thus, the dispersion equation for electromagnetic waves propagating in a bihyrotropic layer is a fourth-order determinant for a system of homogeneous equations (87) with coefficients defined by expressions (88) – (103).

### 8. Calculations of surface spin waves characteristics in ferrite plate

As an example, demonstrating the functionality and usability of the dispersion equation (87), let us consider some characteristics of SW in a ferrite plate.

It is already clear that for SW the main difference between the obtained description and former descriptions in magnetostatic approximation [11] and without it [16, 23], is that the wave distribution inside ferrite plate along the *x*-coordinate is described by two wave numbers[7] – $k_{x21}$ и $k_{x22}$!

Let us demonstrate how this difference influences on the characteristics of the surface SW. Isofrequency dependences for surface SW with different frequencies are presented in Fig. 3, where red curves *1 – 4* are calculated according with presented theory, and black curves *1' – 4'* are calculated in magnetostatic approximation. The calculations were carried out at the following parameters: $H_0 = 300$ Oe, $4\pi M_0 = 1750$ Gs, $s = 40$ μm.

Fig. 3 shows that the isofrequency curves *1 – 4* and the corresponding curves *1' - 4'* differ from each other only at frequencies close to the value $f_\perp = \omega_\perp / 2\pi = \sqrt{\omega_H^2 + \omega_H \omega_M} / 2\pi = 2197.7$ MHz. However, if we look how the wave numbers $k_{x21}$ and $k_{x22}$ characterizing the microwave field distribution over the ferrite plate thickness vary along the iso-frequency curves and compares this variation with a similar variation of the wave number $k_{x2ms}$ calculated in magnetostatic approximation [11], we can see significant differences (see Fig.4).

As can be seen from Fig. 4, at angles φ close to the cut-off angles of wave vector, the magnetostatic dependences $k_{x2ms}(\varphi)$ pass near the curves $k_{x22}(\varphi)$, while at φ = 0 these dependences pass near the curves $k_{x21}(\varphi)$, and the difference in the values $k_{x22}(\varphi = 0)$ and $k_{x21}(\varphi = 0)$ depends significantly on frequency, changing from 255 cm$^{-1}$ at $f = 2198$ MHz to 2 cm$^{-1}$ at $f = 2500$ MHz.

---

[7] This is agree with earlier results for backward SW propagating along the direction of vector H0 in a tangentially magnetized ferrite plate [24].

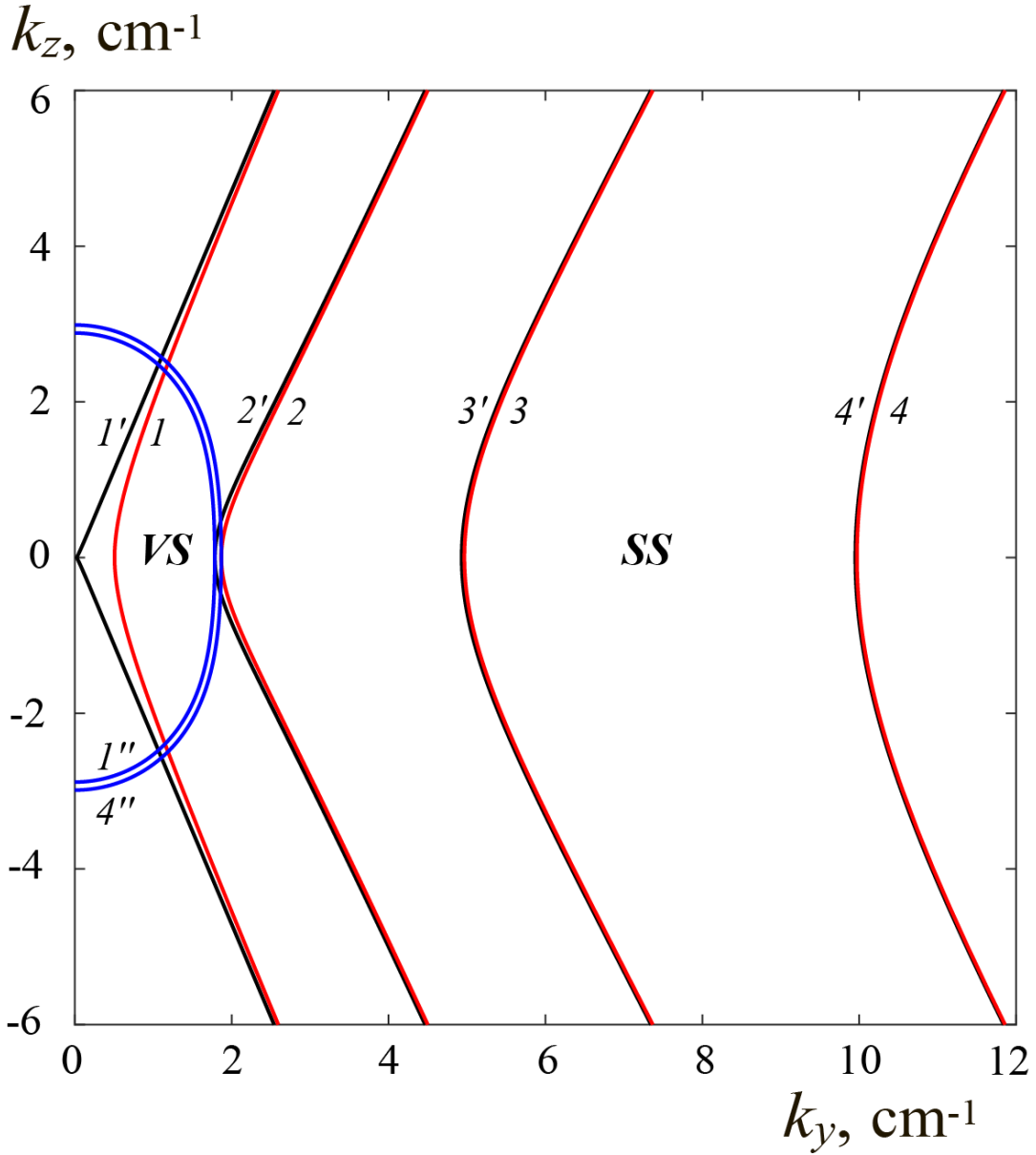

Fig. 3. Isofrequency dependences of the surface SW in the tangentially magnetized ferrite plate for frequencies 2198 (*1* and *1'*), 2216.3 (*2* and *2'*), 2250 (*3* and *3'*) and 2300 MHz (*4* and *4'*) (only half-plane for $k_y > 0$ is shown). Curves *1' – 4'* (black) are calculated in magnetostatic approximation and curves *1 – 4* (red) – without this approximation. Here are also shown curves *1"* and *4"*, which are the intersection of surface $\alpha = 0$ and planes $f = 2198$ MHz and $f = 2300$ MHz, respectively (curve *1"* separates on curve *1* the region with a SS-waves and the region with VS-waves for $f = 2198$ MHz, while curve *4"* does not intersect curve *4*).

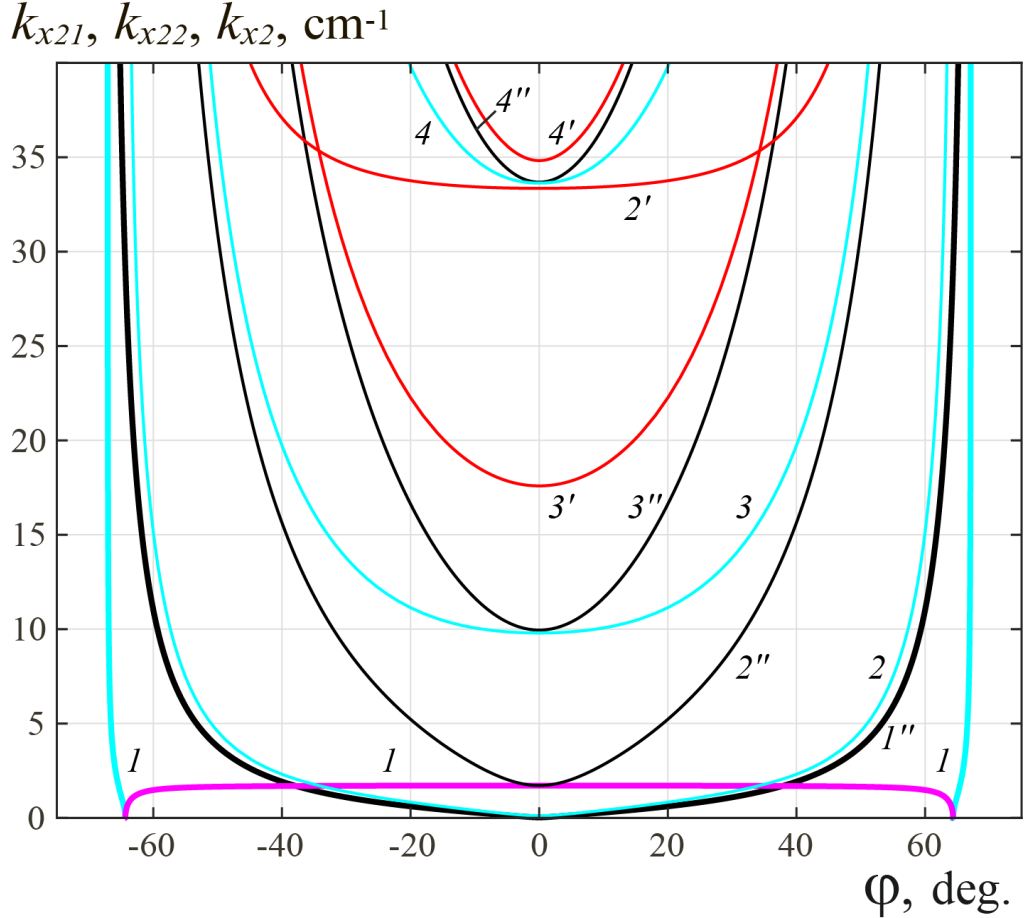

Fig. 4. Dependences of the wave numbers $k_{x21}$ (green curves $1 - 4$), $k_{x22}$ (red curves $1' - 4'$) and $k_{x2ms}$ (black curves $1'' - 4''$) for frequencies 2198, 2216.3, 2300 and 2500 MHz respectively on the angle $\varphi$ indicating the wave vector orientation. Curve $1'$ is not shown in the figure because it is located above of ~ 250 cm$^{-1}$. In the part of curve $1$ shown in purple, $k_{x21}$ takes on imaginary values corresponding to the VS-wave (the value $|k_{x21}|$ is shown in the ordinate axis).

## 9. Comparison of surface spin wave propagation along *y*-axis with results obtained earlier

It is well known that description (without magnetostatic approximation) of the surface SW propagating in a tangentially magnetized ferrite plate perpendicular to magnetic field vector $\mathbf{H_0}$ was obtained many years ago (see, for example, [17, 18]). For this case the Maxwell equations are known to be solved quite easily and the resulting distribution of surface SW over the ferrite plate thickness is characterized by one single wave number but not two. Thus, at first sight one can see a clear contradiction between the above description of surface SV and the description obtained earlier. In fact, as will be shown below, this contradiction is only seeming.

Obviously, the case of SW propagation perpendicularly to the vector $\mathbf{H_0}$ is obtained from the general case of wave propagation in arbitrary direction if we put $k_z = 0$ (or $\varphi = 0$) in the above consideration. In this case (when $k_z = 0$), the system of Maxwell equations (4) splits into two independent subsystems, one of which includes only field components $E_{z2}$, $H_{x2}$ and $H_{y2}$, and the other one – $H_{z2}$, $E_{x2}$ and $E_{y2}$. Since, when $k_z = 0$ from (11) we obtain $F_{vg} = 0$, the resulting equations system (8) for amplitudes $e_{z2}$ and $h_{z2}$ turns into two independent homogeneous Helmholtz equations, one for the amplitude $e_{z2}$ and the other for the amplitude $h_{z2}$. The first equation describes the known surface SW (*H*-wave with components $E_{z2}$, $H_{x2}$ and $H_{y2}$) for which $k_{x22} = k_0\sqrt{F_v} = \sqrt{k_y^2 - k_0^2 \varepsilon \mu_\perp}$ (this value of $k_{x22}$ is the result of substituting of equality $F_{vg} = 0$ into expression (21)), and the second equation describes the surface electromagnetic wave (*E*-wave with components $H_{z2}$, $E_{x2}$ and $E_{y2}$, typical for a layer of usual dielectric) for which $k_{x21} = k_0\sqrt{F_g} = \sqrt{k_y^2 - k_0^2 \varepsilon \mu_{zz}}$ (this value of $k_{x21}$ is the result of substituting of equality $F_{vg} = 0$ into expression (20)). As can be seen, the mentioned values $k_{x22}$ and $k_{x21}$ correspond to results previously obtained (see, for example, [13, 17, 29]).

However, the reader who has carefully studied Figure 4 may see a discrepancy with the above arguments and say: but how so – it is clear that, according to the calculations in Fig. 4 the surface SW is characterized not by a single wave number at $\varphi = 0$ and $k_z = 0$ (as it was established earlier), but by two wave numbers $k_{x21}$ and $k_{x22}$?!

To answer this observation, the change of coefficients *A*, *B* and *C*, normalized to coefficient *D*, as a function of the wave number $k_z$, is calculated below (Fig. 5). Remind here that coefficients *A*, *B*, *C*, and *D* define amplitudes of exponential functions in expression (26). And as we can see from Fig. 5, the coefficients $A$ and $B$ near the exponents $\exp(k_{x21}x)$ and $\exp(-k_{x21}x)$ become zero at $k_z = 0$ (or $\varphi = 0$), while the coefficients $C$ and $D$ near the exponents $\exp(k_{x22}x)$ and $\exp(-k_{x22}x)$ do not become zero! Thus, it is clear that at $k_z = 0$ (or $\varphi = 0$) the surface SW distribution over the ferrite plate thickness is described by a single wave number $k_{x22}$.

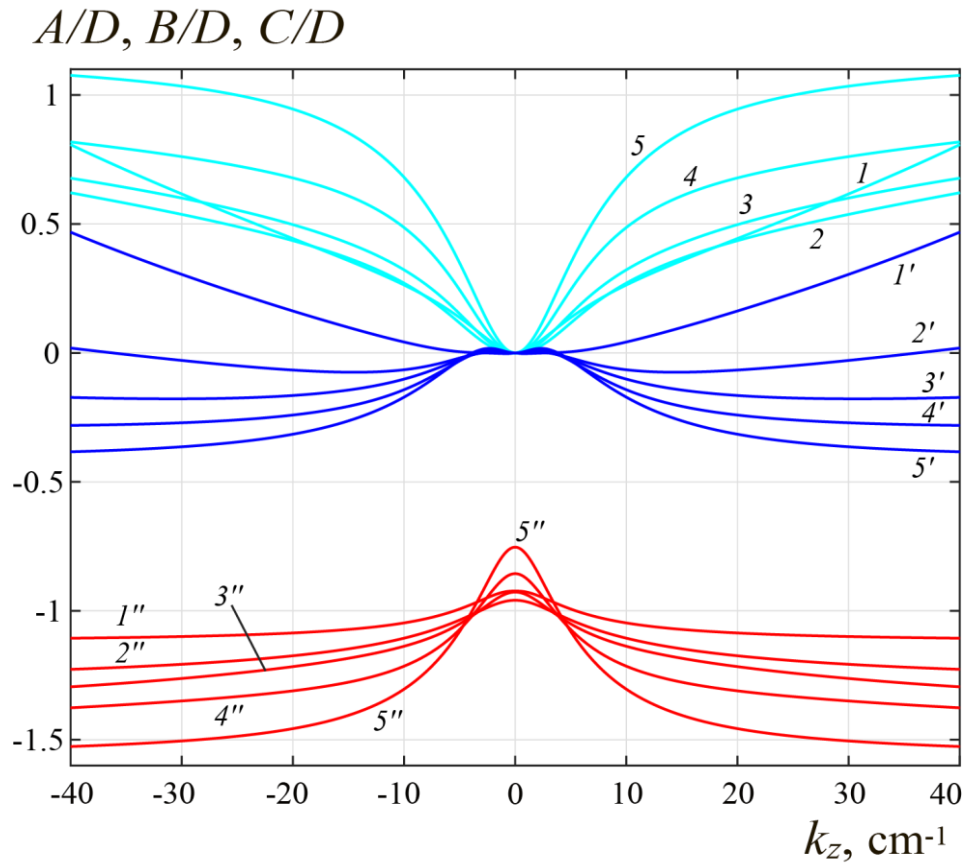

Fig. 5. Dependencies of coefficients ratio $A/D$ (light-blue curves $1 – 5$), $B/D$ (blue curves $1' – 5'$) and $C/D$ (red curves $1'' – 5'''$) on the wave number $k_z$ for frequencies 2217, 2300, 2500, 2800 and 3200 MHz respectively.

Basing on the data of Fig. 4, we note that when $\varphi \sim 0$ and $k_z \sim 0$, the dependence $k_{x21}(\varphi)$ does not really describe the SW distribution over the ferrite plate thickness, since, as can be seen from Fig. 5, the coefficients $A$ and $B$ for all frequencies are practically equal to zero at $k_z \sim 0$. Thus, in accordance with formulas (55) – (64), the main contribution to the microwave field amplitude distribution at $\varphi \sim 0$ and $k_z \sim 0$ is provided by the coefficients $C$ and $D$ and their corresponding wave number $k_{x22}$. Therefore, it should be taken into account that description of the surface SW distribution over the ferrite thickness in magnetostatic approximation is rather imprecise for the initial part of the SW spectrum (in the range ~ 200 MHz above the frequency $f_\perp$) at $\varphi \sim 0$ and $k_z \sim 0$, where $k_{x2ms}(\varphi)$ dependences almost coincide with $k_{x21}(\varphi)$ dependences, but lie quite far from $k_{x22}(\varphi)$ dependences.

## 10. Conclusion

Analytically, without magnetostatic approximation, there is solved the general problem of electromagnetic wave propagation along arbitrary direction in tangentially magnetized plane-parallel bigyrotropic layer described by the dielectric and magnetic permittivities in the form of Hermitian tensors of second rank. It is shown that by presenting a solution of the Maxwell equations in the layer's plane as the wave of form $\sim \exp(-ik_y y - ik_z z)$ and by leaving an arbitrary the wave dependence on *x*-coordinate (that is normal to the layer's plane), one can bring the Maxwell equations to a system of two differential equations containing only *x*-dependent amplitudes of the microwave electric and magnetic fields, parallel to the vector of constant homogeneous magnetic field **H₀**. In its turn, this system is reduced to a fourth order linear differential equation, to which corresponds the biquadratic characteristic equation determining the wave numbers of electromagnetic wave distribution over the layer's thickness. It is shown that this characteristic equation has four simple (non-multiple) roots $k_{x21}$, $k_{x22}$, $k_{x23} = -k_{x21}$ and $k_{x24} = -k_{x22}$, which cannot be complex numbers and can take only real or imaginary values.

A boundary problem for propagation of electromagnetic waves with real values of $k_{x21}$ and $k_{x22}$ in a bihyrotropic layer surrounded by dielectric half-spaces is solved and the dispersion equation presenting the fourth-order determinant for a system of homogeneous linear equations is derived for these waves. It is shown that the electromagnetic wave propagating in the layer's plane in an arbitrary direction has all six components of the microwave electromagnetic field – three magnetic and three electrical – both in the ferrite layer and in the adjacent half-spaces. Physically, this means that satisfying of the boundary conditions on the layer's surfaces results in appearance of both *E*-wave and *H*-wave in the adjacent half-spaces, and these two waves are interconnected to each other by means of the bihyrotropic layer.

Based on this theory, the characteristics of spin waves in a ferrite plate, which is a special case of a bihyrotropic layer, have been studied. It is shown that in the ferrite plate and in the simplest structures based on it, the wave distribution inside ferrite plate can be described by both exponential and trigonometric functions corresponding to the surface and volume wave distributions. Thus, on the waves dispersion surfaces $f(k_y, k_z)$ corresponding to these structures, there can be areas describing three types of propagating waves with different distributions over the ferrite thickness: surface-surface waves ($k_{x21}$ and $k_{x22}$ are real numbers), volume-surface waves ($k_{x21}$ is imaginary and $k_{x22}$ is real) and volume-volume waves ($k_{x21}$ and $k_{x22}$ are imaginary numbers). The fourth case of the surface-volume wave (where $k_{x21}$ is real number and $k_{x22}$ is imaginary number) is not realised in the ferrite plate and in structures based on it. In the $\{k_y, k_z, f\}$ coordinate space we graph the boundary surfaces that, by crossing a certain dispersion surface of spin waves, would separate on it areas with different wave distributions. It is shown that the boundary surfaces are described by equation identical to the dispersion equation for electromagnetic waves in unbounded ferrite (bihyrotropic) medium if the latter would be simplified to the two-dimensional case.

The characteristics of surface SW calculated in magnetostatic approximation have been compared with similar characteristics calculated without it. It is found that the isofrequency dependences calculated by both methods differ appreciably only in the region of small wave numbers $k < \sim 3$ cm$^{-1}$ for a frequency interval $\sim 20$ MHz lying above the frequency $f_\perp$. It is shown that coefficients *A* and *B* near the exponents $\exp(k_{x21}x)$ and $\exp(-k_{x21}x)$ in SW distribution over the ferrite thickness become equal to zero at $k_z = 0$ (or $\varphi = 0$), so the wave distribution in this case is described by the single wave number $k_{x22}$. It is found that dependences of the wave numbers $k_{x21}$ and $k_{x22}$ on the orientation $\varphi$ of the wave vector for the surface SW are significantly different from the analogous magnetostatic dependence $k_{x2ms}(\varphi)$ in the wide frequency band ~ 200 MHz lying above the frequency $f_\perp$. In particular, at angles φ close to the wave vector cut-off angles, the $k_{x2ms}(\varphi)$ dependences pass near the $k_{x22}(\varphi)$ curves and at φ ~ 0 near the $k_{x21}(\varphi)$ curves, and the difference in the $k_{x22}(\varphi = 0)$ and $k_{x21}(\varphi = 0)$ values changes significantly with frequency from ~ 255 cm$^{-1}$ at $f \sim f_\perp$ to ~ 2 cm$^{-1}$ at $f \sim f_\perp + 300$ MHz. Therefore, the magnetostatic description of surface SW distribution over the ferrite thickness is rather imprecise for the initial part of surface SW spectrum at φ ~ 0 and $k_z$ ~ 0, where the $k_{x2ms}(\varphi)$ dependences lie far from the $k_{x22}(\varphi)$ dependences and almost coincide with the $k_{x21}(\varphi)$ dependences, really not describing the wave (since $A(\varphi = 0) = B(\varphi = 0) = 0$).

The obtained solution of the problem on propagation of electromagnetic waves in a tangentially magnetized bihyrotropic layer opens wide possibilities for calculation of exact wave's characteristics in layers of such media as ferrite, antiferromagnetic and plasma. In particular, not only the Poynting vector, direction and density of energy flux and vector lines for the studied waves can be calculated by means of the presented theory but also more complicated problems can be solved on the basis of available electrodynamic methods.

## Funding